# Who Owns the Data? A Systematic Review at the Boundary of Information Systems and Marketing

Stephen L. France
College of Business
Mississippi State University
Email: sfrance@business.msstate.edu

Mahyar Sharif Vaghefi
College of Business Administration
University of Texas at Arlington
Email: Mahyar.sharifvaghefi@uta.edu

Brett Kazandjian
College of Business and Economics
Towson University
Email: bkazandjian@towson.edu

**Abstract**

This paper gives a systematic research review at the boundary of the information systems (IS) and marketing disciplines. First, a historical overview of these disciplines is given to put the review into context. This is followed by a bibliographic analysis to select articles at the boundary of IS and marketing. Text analysis is then performed on the selected articles to group them into homogeneous research clusters, which are refined by selecting "distinct" articles that best represent the clusters. The citation asymmetries between IS and marketing are noted and an overall conceptual model is created that describes the "areas of collaboration" between IS and marketing. Forward looking suggestions are made on how academic researchers can better interface with industry and how academic research at the boundary of IS and marketing can be further developed.

# Introduction

The current era of marketing is ruled by data. Marketers need to process large amounts of data from a myriad of sources including corporate databases, weblogs, online reviews sites, social media platforms, and smartphone check-ins in order to engage with customers and plan marketing campaigns. However, data-based marketing is not new. As early as the 1960s, several articles noticed the need for the efficient use of data and for the use of information systems (IS[1]) in marketing. Brien and Stafford (1968) noted that "marketing is inextricably caught up in the communications revolution. The new era, the age of information, will emphasize the information gathering and processing structure of the organization." This integration of marketing and information was to be achieved using marketing information systems. For example, Adler (1967) posited that traditional marketing research was becoming obsolete and was to be replaced by marketing information systems that are "tailored to the needs of each marketer". Berenson (1969) noted that these systems were needed due to the increased complexity of business processes, increasing volumes of data, shorter product lifecycles, and the development of sophisticated data analytic techniques. All these arguments still hold true over 50 years later.

This review aims to shed light on some of the developments that have occurred between the initial awareness and implementation of marketing information systems in the 1960s and the current era of "big data" and sophisticated analytics. The review operates at the boundary of the marketing and IS disciplines and examines areas of overlap and collaboration between the disciplines in the context of a modern e-commerce marketing technology environment.

The review is a systematic review in that it utilizes an "organized, transparent, and replicable process" (Littell et al., 2008) to gather source material for the review and utilizes text-based content analysis methods to classify the findings and make conclusions (Paré et al., 2015). Content analysis methods for selecting and classifying articles for a systematic review can include text mining, citation analysis, and topic modeling (Barczak, 2017). A citation analysis strategy, using text-mining and analysis, was used to match journal names. Articles were selected from leading marketing and IS journals and were included in the review if they had more than a certain percentage of cross citations between marketing and IS. These cross-cited articles were used to define the work at the boundary of marketing and IS and to extract research trends over time.

---

[1] We use the acronym IS for the remainder of the paper, as Information Systems is the most common discipline name and used by the leading societies in the discipline, but note that other acronyms, such as, IT, MIS, and ITM, are also used.

The remainder of the paper starts with a description of early work in marketing information systems, with respect to the development of the marketing and IS disciplines. A short history of each discipline is given, along with a discussion of the major journal outlets in each discipline. This information is then used to build a systematic review at the intersection of the two disciplines. Citation analysis and text analysis of paper abstracts are used to find areas of collaboration between the disciplines. These "areas of collaboration" are analyzed thoroughly, and the results are used to build a conceptual model of collaboration between marketing and IS.

*The Marketing Discipline*

The academic marketing discipline has been around since the first half of the twentieth century, and its maturation and growth were tied in to the general growth in business school education in this period, spurred by the growth of the U.S. economy, the rapid increase in higher education enrollment, and the opening of the first dedicated business schools, including business schools at the University of Pennsylvania in 1881 (Sass, 1982), the University of Chicago in 1898, and Harvard in 1908 (Cruikshank, 1987). Wilkie and Moore (2003) divided the development of academic research in marketing into four different eras. In the founding era of 1900-1920, the first marketing courses were developed, and marketing had a rather narrow focus as a distribution activity. In the formalization era of 1920-1950, academic infrastructure was built up, including conferences and pioneering journals, such as the Journal of Marketing and the Journal of Retailing.

In the era of 1980-present, there has been increasing fragmentation of marketing thought into sub-disciplines and growth in the knowledge infrastructure of associations, conferences, and journals. A survey of marketing citations (Baumgartner & Pieters, 2003) noted 49 marketing journals split into five areas. A "meta-list" of business journals (Harzing, 2019) lists 67 marketing journals.

Despite the rapid increase in the number of journals, the top journals in marketing, as shown in Table 1, have remained relatively constant. Here, a top journal is one that is included in at least one major list of "top journals" (Bloomberg, UT-Dallas, FT), has the top (4*) ranking on the Chartered Association of Business Schools academic journal guide (CABS, 2018), and is highly ranked by the widely cited journal perceptions article by Theoharakis and Hirst (2002). The Journal of Marketing Research was spun off from The Journal of Marketing and had an initial focus on research methods (Lehman, 2005). The Journal of Consumer Research was founded to provide an outlet for the growing behavioral research community (Frank, 1996), and the Journal of the Academy of Marketing Science was established as a general interest journal in 1973 to support the newly formed Academy of Marketing Science (Malhotra, 1996). The most recent of these journals is Marketing Science, which was founded in 1982 by the ORSA/TIMS College on Marketing as an outlet for quantitative marketing work (Little, 2001; Winer & Neslin, 2014; Wittink, 2001).

**Table 1. Top-Tier Marketing Journals**

| Journal | Founded |
|---|---|
| Journal of Marketing (JM) | 1936 |
| Journal of Marketing Research (JMR) | 1964 |
| Journal of the Academy of Marketing Science (JAMS) | 1973 |
| Journal of Consumer Research (JCR) | 1974 |

The last two eras, also known as the "paradigm shift" and "paradigm broadening" eras (Shaw & Jones, 2005), are periods where research in marketing information systems first materialized. This stream of research arose as a result of an increased emphasis on and utilization of quantitative methods in business, as well as an increased interest among management academics and practitioners in the use of computers in management analysis (Coleman, 1956; Hurd, 1955). Much of this early work came under the auspices of "management science" or "operations research". The Management Science journal was founded in 1954 and within a few years began to publish marketing-related articles, for example, a managerial overview of the applications of management science in marketing (Anshen, 1956).

In addition, there was much crossover work designed to bring operations research methods to a marketing audience, including two articles in the Journal of Marketing (Doherty, 1963; Magee, 1954). This management science tradition in marketing was supported by the marketing college of TIMS (Morrison & Ragu, 2004), which ultimately led to the Marketing Science journal and conferences (Wittink, 2001).

Aside from Adler (1967) and Berenson (1969), there was a stream of systems work emerging from the marketing and management science areas on how to utilize a range of quantitative methods, such as forecasting, mathematical programming, and simulation, in information systems for marketing planning (e.g., Casher, 1970; Montgomery & Urban, 1969; Montgomery & Urban, 1970). In this era, some initial thought was given to managerial aspects of quantitative methods; in particular, decision calculus (Little, 1970; Chakravarti et al., 1979), which provides a framework for integrating quantitative models into managerial marketing processes. A range of modeling tools for marketing systems followed this initial work, including computational tools for test marketing (Urban, 1970), marketing mix optimization (Little, 1975; Lilien, 1979), and product positioning (Urban, 1975). A discussion of how these marketing "decision support systems" are used in practice is given in Little (1979), who noted that these systems are at the boundary of marketing and IS, and that their success is reliant on IS infrastructure and internal IS resources, which may sometimes create conflict. This reliance on IS resources helped lead to more general work in decision support systems in the nascent discipline of IS (Keen, 1980a,b) and IS work on developing marketing systems (e.g., Lee & Lee, 1999; Raghu et al, 2001).

*The Management Information Systems Discipline*

Compared to marketing, the IS discipline was formalized and developed much later. The first formal academic program in management information systems was developed in 1968 at the University of Minnesota (Nolan & Wetherbe, 1980), which coincided with a growth in computer installations from 15,000 to 85,000 between 1963 and 1970 (Dickson, 1981). Early work in IS, which was predominantly focused on systems, helped apprise and make sense of this new environment. For example, Aron (1969) gave a taxonomy of data processing, information retrieval, and management information systems and Orlicky (1969) gave a blueprint for planning and developing computer systems. Dickson (1981) provided a detailed summary of this early work.

Much of the early work in IS was considered to be part of the general quantitative management literature and was published in outlets such as Management Science, which had specific departments for IS as early as 1969 (Banker & Kauffman, 2004). This work built on early work on the use of computers in management (Coleman, 1956) and general systems theory (Boulding, 1956). It includes work on IS assumptions and deficiencies (Ackoff, 1967), a behavioral and organizational taxonomy for IS (Mason & Mitroff, 1973), and the role of managerial involvement in IS success (Swanson, 1974). In addition, there was much general work on management systems, for example, Raymond (1966) and Beged-Dov (1967).

As with marketing, a knowledge infrastructure, including societies, conferences, and journals, grew to accommodate the nascent IS discipline (Hirschheim & Klein, 2012). A list of the four leading journals in the field is given in Table 2. A similar methodology was used as the marketing list, but journals ranked (4) on the ABS list were allowed, as only two journals (Information Systems Research and MIS Quarterly) had 4* rankings.

**Table 2. Top-Tier IS Journals**

| Journal | Founded |
|---|---|
| Management Information Systems Quarterly (MISQ) | 1977 |
| Journal of Management Information Systems | 1984 |
| Information Systems Research (ISR) | 1990 |
| Journal of the Association of Information Systems | 2000 |

MIS Quarterly was founded in 1977 to provide a dedicated outlet for IS researchers and was intended to serve both academic researchers and practitioners (Rai, 2016). Shortly after, the first major IS conference, the International Conference on IS, was launched in 1980 (King & Galletta, 2010). The IS discipline was further enriched in the 1980s by a range of workshops sponsored by the Harvard Business School covering research areas such as qualitative, experimental, and survey research (Benbasat & Weber, 1996). The Journal of Management Information Systems (JMIS) was founded in 1984 with the goal of providing an "integrated view" of the maturing IS discipline (Zwass, 1984). Information Systems Research, published under the same TIMS (now INFORMS) umbrella as Management Science, followed in 1990, again with a mission to publish research representative of all areas of IS and "address the application of information technology to human organizations and their management" (Swanson et al., 2010). The Journal of the Association of Information Systems was introduced in 2000 as the society journal of the Association of Information Systems, which was started as an association in 1995 and has since grown to become a center point of the IS field (King & Galletta, 2010) and a truly global organization, with conferences across the world.

*Summary and Rationale for Review*

The previous sections gave summaries of the early histories of the marketing and IS disciplines. The quantitative marketing and IS disciplines have several similarities. Both have roots in general management science, had a focus on decision support systems in their early years, and have had a knowledge infrastructure of journals and conferences build up over time. While marketing is an older, more mature discipline than IS, the growth of decision support systems research in IS coincided with the growth of research in marketing models and systems.

But, in the current data driven management world, what is the relative positioning of the two different disciplines? In the practitioner world, there are many areas that are at the boundary of information systems and marketing. Examples include customer relationship management (Buttle, 2004), social media analytics (Fan & Gordon, 2014), search engine optimization (Davis, 2006), and location analytics (Garber, 2013). The creation of dedicated analytics teams and the employment of "data scientists" have blurred the boundaries between the disciplines. For example, at the data driven web company Quora, around half of product managers have a data science background (Mayefsky, 2018). In addition, modern marketing platforms incorporate data and computational technologies, such as artificial intelligence and machine learning (Olson & Levy, 2018). However, given disciplinary traditions, it may be that the marketing and IS disciplines look at these new developments from very different perspectives. Accordingly, the major aims of this review are as follows. First, to understand how marketing and IS take different viewpoints of practitioner problems at the boundary of marketing and information systems. Second, to see how closely marketing and IS research overlap. Is there research that could safely fit in either field or is there a gulf between the fields? Is theory built across the two disciplines, or are the theoretical constructs for the disciplines quite separate? Third, to see how major technological innovations over the last 20 years, such as e-commerce and social media, have been treated differently by the two fields, and to propose a pathway for research at the boundary of marketing and IS that incorporates knowledge and innovations from both disciplines. Forth, to give a set of forward-looking recommendations for both academic work and academic/practitioner collaboration at the boundary of the marketing and IS disciplines.

## The Bibliographic Analysis and Review Process

To seed the review process, citations were collected from the previously selected lists of top marketing and IS journals. In addition, the Management Science journal was included as a reference journal, given the previously described importance of the journal in the history of both the quantitative marketing and IS areas.

The period of the review is from 1990-2017. As previously described, the knowledge infrastructure for IS was built up much earlier than the knowledge infrastructure for marketing. The discipline was still maturing in the 1970s and 1980s, with the currently leading journals and conferences still being defined. The year 1990 was chosen for the start of the review, as by the beginning of that year, three out of the four selected top-tier IS journals had been started. In addition, the period of 1990-2017 encompasses major disruptive technological events, such as the growth of the internet and the fragmentation of the computer market caused by smartphones and tablets.

The systematic review process was implemented by building the review around selected papers that are at the boundary of marketing and information systems. Citation analysis was used to select these papers, as it is a powerful tool for gaining insight into research patterns (Meho, 2007) and is a widely used method in management research (e.g., Eppler & Mengis, 2004; Klang et al., 2014; Liñán & Fayolle, 2015). Abstract information for the selected articles, combined with text analysis techniques and document clustering, was used to help frame the terms of reference for the review.

*The Research Process*

A web crawler, written in Python, was used to download information from the Web of Science API for each article from each of the selected journals from 1990-2017. The Web of Science is a detailed journal and citation repository (Mongeon & Paul-Hus, 2016) and has been used to gather data for bibliometric analyses and reviews in business, with applications including an analysis of a research sub-field, for example, social entrepreneurship (Rey-Martí et al., 2016) or international marketing (Samiee & Chabowski, 2012), or an examination of the positioning of an individual journal (e.g., Merigó et al., 2015).

For each article, the author, title, abstract, and author information was retrieved along with all the citations. For each citation, the journal name (if a journal article) was recorded, along with the Web of Science journal category. The Web of Science journal categories are somewhat broad (for example "Business & Management") and do not distinguish between the different business disciplines. To subcategorize the business journals, categories were utilized from the Harzing list (2019), which incorporates information from twelve different journal quality and categorization studies. For the sake of parsimony, the non-business journals were grouped together under a single category "other", though computer science was kept separate due the linkage and shared sub-disciplines (e.g., software engineering) between information systems and computer science. The final categories are Computer Science, Economics, Finance & Accounting, Information Systems, Management, Marketing, None, Operations and Supply Chain, and Other. The "None" category includes non-Web of Science journal publications, such as conference proceedings, working papers, and theses. Several of the journals (JAMS, JAIS, and JMIS), did not have Web of Science information for the first few years of the search in the 1990s, so for each of these journals, individual web crawlers and parsers were written to retrieve journal and citation information from the journal webpages and the resulting data were combined with the Web of Science data.

*Exploratory Analysis*

In total, 684,196 references were gathered from 14,282 journal articles. Of these references, 62% came from Web of Science listed journals and 38% came from conference proceedings, working papers, theses, and unlisted papers. All Web of Science cites were categorized. The results are summarized by percentage for the Web of Science journals in Table 3. Overall, around 87.5% of Web of Science cites were to business journals, with 12.5% of cites going to other journals. JCR had a far higher percentage of cites going to "other" journals than any other selected journal (46.9%). This is likely to be because JCR sits at the boundary of marketing and the behavioral sciences and thus sends a large number of cites to disciplines such as psychology and sociology.

The marketing cites are more endogenous than the IS cites, with marketing cites to marketing journals ranging from 45.37% (JCR) to 55.74% (JAMS) and IS cites to IS journals ranging from 30.98% (ISR) to 45.08% (JAIS). The citation patterns for Management Science journal fit with its positioning as a general quantitative methods journal, with 24.97% of cites going to economics, 23.51% of cites going to operations, and significant cites going to finance & accounting (12.71%), management (11.51%), and marketing (7.01%).

**Table 3. Journal Citation Percentage Category Summary (Web of Science only)**

|  | Categories | | | | | | | |
|---|---|---|---|---|---|---|---|---|
| **Journal** | **Comp Sci.** | **Econ** | **Fin. & Acc.** | **MIS** | **Management** | **Mkt.** | **OM & SCM** | **Other Journal** |
| Information Systems Research (ISR) | 2.87% | 9.98% | 1.67% | 30.98% | 19.39% | 8.34% | 10.61% | 16.16% |
| Journal of Consumer Research (JCR) | 0.14% | 2.70% | 0.29% | 0.18% | 3.71% | 45.37% | 0.71% | 46.90% |

| Journal | | | | | | | | |
|---|---|---|---|---|---|---|---|---|
| Journal of Management Information Systems (JMIS) | 3.82% | 5.47% | 1.57% | 40.25% | 20.05% | 5.79% | 8.72% | 14.33% |
| Journal of Marketing (JM) | 0.14% | 5.71% | 1.80% | 0.77% | 17.71% | 54.18% | 3.18% | 16.53% |
| Journal of Marketing Research (JMR) | 0.30% | 9.21% | 1.62% | 0.41% | 7.52% | 50.39% | 3.83% | 26.72% |
| Journal of the Academy of Marketing Science (JAMS) | 0.15% | 2.98% | 1.13% | 1.05% | 22.35% | 55.74% | 2.64% | 13.97% |
| Journal of the Association of Information Systems (JAIS) | 3.00% | 2.91% | 1.13% | 45.08% | 19.24% | 4.71% | 6.63% | 17.29% |
| Management Science (ManSci) | 1.00% | 24.97% | 12.71% | 2.90% | 11.51% | 7.01% | 23.51% | 16.38% |
| Marketing Science (MarSci) | 0.52% | 21.25% | 1.22% | 0.81% | 4.85% | 51.22% | 8.13% | 11.98% |
| Management Information Systems Quarterly (MISQ) | 2.12% | 4.88% | 1.64% | 37.50% | 22.70% | 7.06% | 7.71% | 16.38% |

The citation patterns between marketing and IS are rather asymmetric. In order, the marketing journals give 0.29% (JCR), 0.41% (JMR), 0.77% (JM), 0.81% (MarSci), and 1.05% (JAMS) of their journal cites to IS journals, while the IS journals give 4.71% (JAIS), 5.79% (JMIS), 7.06% (MISQ), and 8.34% (ISR) of their journal cites to marketing journals.

The main purpose of the citation analysis was to generate a set of candidate papers at the boundary of the marketing and IS disciplines for a systematic review. Several numerical cut-offs were examined with the goal of selecting a set of papers that was large enough for a serious review, but that could keep the review to a sensible "journal paper length". In the end, papers with at least 10% marketing journal cites and 10% IS journal cites were selected. In addition, a criterion of a minimum of 5 cites for each field was included to prevent research notes/corrections with only a very small number of references being included. This resulted in a candidate list of 331 journal articles. The distribution of the selected journal articles relative to the overall distribution of all the journal articles is given in Table 4. The asymmetry between cites becomes clearer here. Apart from JAMS (14 articles), which is probably the most broadly strategy-focused and practitioner oriented of the major marketing journals (Berkman, 1992; Palmatier, 2018), no marketing journal had more than five articles selected. Conversely, over 80 articles were selected from each of ISR, JMIS, and MISQ. Overall, 8.5% of MIS papers have at least 10% of their cites directed to marketing journals, indicating that IS researchers draw significantly from the marketing discipline.

**Table 4. Summary Information on Selected Papers**

| Journal | Papers | Percentage | Selected Papers | Selected % |
|---|---|---|---|---|
| Information Systems Research (ISR) | 863 | 6.05% | 82 | 24.77% |
| Journal of Consumer Research (JCR) | 1587 | 11.13% | 1 | 0.30% |
| Journal of Management Information Systems (JMIS) | 1104 | 7.74% | 89 | 26.89% |
| Journal of Marketing (JM) | 1180 | 8.28% | 4 | 1.21% |
| Journal of Marketing Research (JMR) | 1690 | 11.85% | 1 | 0.30% |
| Journal of the Academy of Marketing Science (JAMS) | 1256 | 8.81% | 14 | 4.23% |
| Journal of the Association of Information Systems (JAMS) | 466 | 3.27% | 30 | 9.06% |
| Management Science (ManSci) | 3851 | 27.01% | 14 | 4.23% |
| Marketing Science (MktSci) | 1235 | 8.66% | 4 | 1.21% |
| Management Information Systems Quarterly (MISQ) | 1025 | 7.19% | 92 | 27.79% |
| Total | 14257 | 100.00% | 331 | 100.00% |

To examine citation patterns over time, in Figure 1, the total number of papers per year in the source journals was compared with the total number of selected papers per year. To enable a relative comparison over time, each series was plotted on a different axis. It is easy to see that the number of selected "boundary" articles is very low in the 1990s but increases rapidly from 2000. This indicates that while in the 1990s there may have been issues of interest to both marketing and IS disciplines, marketing and IS researchers publishing in the top journals in the field rarely drew upon one another's work. As will be enlightened by the subsequent content analysis, the growth of e-commerce and the importance of this area to both marketing and IS disciplines has begun to bring the disciplines together. In particular, IS researchers have drawn from marketing theory to help examine areas such as e-commerce, social media, and user-generated content.

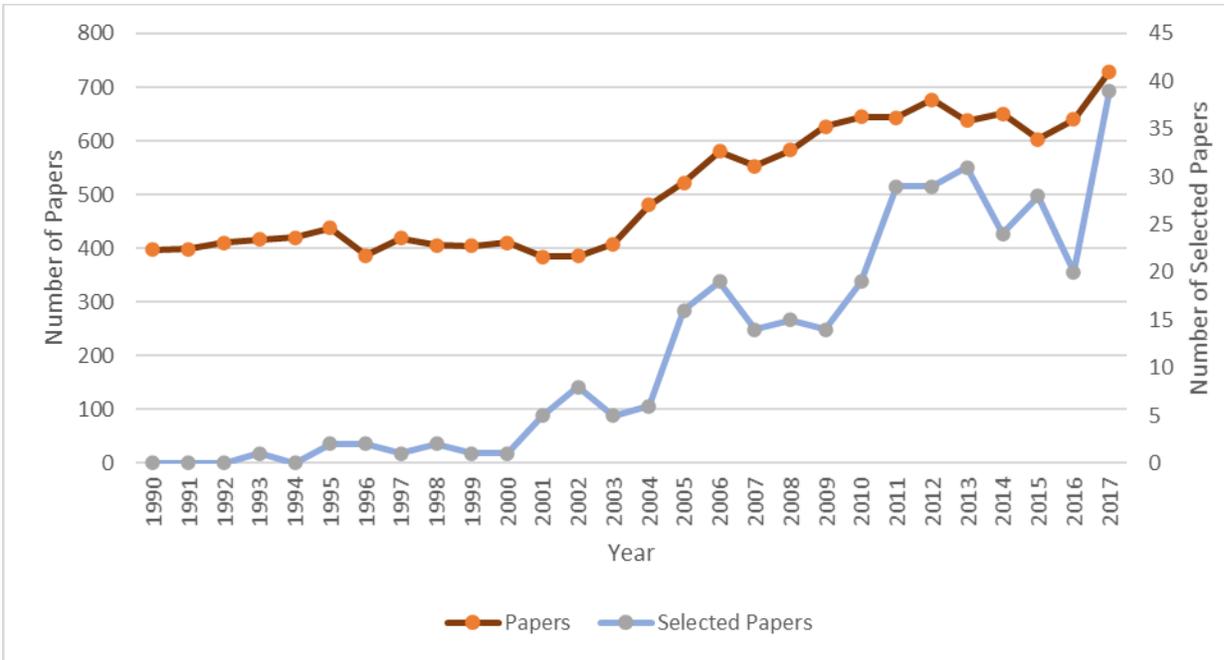

**Figure 1. Comparison of Number of Papers vs. Selected Papers Over Time**

*Content Analysis of Abstracts*

Content analysis is the process of analyzing and coding unstructured data to gain insight and make inferences and find patterns in the data. Content analysis was developed in the field of communications research over the first half of the twentieth century as "a research technique for the objective, systematic and quantitative description of the manifest content of communication" (Berelson, 1952). Content analysis is a method of simplification and is used to code complex representations into simpler categories using a well-defined coding and categorization process (Berelson, 1952; Krippendorff, 1980) to find patterns and overall themes (Hsieh & Shannon, 2005). Content analysis is on the boundary of qualitative and quantitative research, with qualitative analysis of unstructured data often leading to summary data that can be analyzed with quantitative techniques (Morgan, 1993). As early as the 1960s, attempts were made to perform content analysis with computers (Stone et al., 1966), and content analysis has increasingly become more "mechanized", with automated text mining and analysis tools used to code and categorize information (Hopkins & King, 2010; Rajman, & Besançon, 1998; Yu et al., 2011). Automated content analysis has been used to analyze journal article information and abstracts in order to motivate and "seed" systematic academic literature reviews (Delen, & Crossland, 2008).

Examples of such analyses abound from areas including business intelligence (Moro et al., 2015), technology acceptance (Mortenson & Vidgen, 2016), accounting (Chakraborty et al., 2014), marketing personalization (Sunikka & Bragge, 2012), ecology (Nunez-Mir et al., 2016), and science education research (Chang et al., 2010).

To perform the content analysis, the abstracts were combined and preprocessed to remove numbers, punctuation, and a standard list of stop words (common words with low content information). Each of the remaining words was considered a "term" along with each bigram combination of two words. Adding bigrams increases the complexity and dimensionality of the data, but can help give improved text classification/clustering performance (e.g., Koster & Seutter, 2003; Tan et al., 2002), particularly in domains with limited lexicons (Bekkerman & Allan, 2004). The frequency of each term in each document (abstract) was recorded in a document $\times$ term frequency matrix. Each column of the matrix was multiplied by the inverse document frequency (idf) of its term. This so called tf-idf formulation (Roelleke & Wang, 2008) is used extensively in document clustering. Weighting by inverse document frequency gives a measure of how far above a baseline frequency a term is and can be thought of as a measure of relevancy for the term in a specific document (Ramos, 2003; Stephen, 2004; Wu et al., 2008).

The main rationale behind our use of content analysis is to gain insight into the different research areas or "areas of collaboration" that straggle marketing and information systems and to split the 331 selected papers into manageable homogenous subsets that correspond to distinct research areas. Several methods of splitting (or clustering) the documents were tested, including hierarchical cluster analysis, k-medoid clustering, and latent Dirichlet allocation. The best results in terms of cluster homogeneity and face validity were given by first calculating cosine distances between documents and then employing k-medoids based clustering, which is a partitioning clustering technique that is relatively robust to outliers (Jin & Han, 2017). Cosine distances have been found to give strong performance in document clustering applications, particularly when items are being clustered from document/term matrices (France et al., 2012; Muflikhah & Baharudin, 2009; Strehl et al., 2000), as unlike Euclidean distances, cosine distances are invariant with respect to document length.

The k-medoids method does not determine an optimal number of clusters. The number of clusters, k, is given as an input parameter to the technique. There are multiple for determining the optimal number of clusters for partitioning clustering (e.g., Jain, 2010; Steinley, 2016). However, these methods often do not agree with one another. In picking the number of clusters we looked for a solution that had a high degree of face validity and was stable with respect to changes in the data. Solution stability was tested using 10-fold cross validation (Arlot & Celisse, 2010; Kohavi, 1995). The dataset was split into 10 equal subsets each containing 10% of the data, and for each subset, clusters were created using the remaining 90% of the data (the training dataset). The remaining 10% holdout data were added to the cluster with the nearest cluster centroid (center) created using the training data. Each of the 10 solutions was then compared for similarity with each of the other solutions using the Hubert-Arabie adjusted Rand index (Hubert & Arabie, 1984), which measures cluster agreement and accounts for random error. An aggregate value of the index was calculated across all 45 = (10(10-1)/2) pairs of clustering solutions. We found that six to eight cluster solutions gave good face validity with respect to homogeneous research areas. Of these solutions, the seven cluster solution had a global maxima for the adjusted Rand index validation measure, so this solution was chosen to seed the review.

To help understand the clusters, we calculated the lift for each term in each cluster. Lift was defined in (1) for term $j$ and cluster $k$ as the increase in the proportion of the term in the tf matrix entries for documents in cluster $k$ over its proportion across all the documents.

$$Lift(j,k) = \frac{\sum_{i \in C_k} tf_{ij}}{\sum_{i \in C_k} \sum_{l=1}^{m} tf_{il}} - \frac{\sum_{i=1}^{n} tf_{il}}{\sum_{i=1}^{n} \sum_{l=1}^{m} tf_{il}} \qquad (1)$$

, where $tf_{ij}$ is the term frequency for term $j$ in document $i$, $m$ is the number of terms, and $n$ is the number of documents. Table 5 gives the terms with the top five lift values for each of the features. Most of the terms are single words, but several bigrams are present, including service_quality, social_media, supply_chain, and information_privacy.

**Table 5. Cluster Top 5 Terms**

| Cluster | Name | No. Articles | Top 5 Terms |
|---|---|---|---|
| 1 | Consumer Trust | 41 | trust, usage, model, web, satisfaction |
| 2 | Consumer Decision Making | 94 | decision, consumers, product, purchase, shopping |
| 3 | Services & Loyalty | 53 | service, quality, service_quality, customer, services |
| 4 | Recommendation Systems | 43 | reviews, product, review, online, products |

| 5 | Word of Mouth | 50 | social, firm, media, social_media, customer |
| 6 | Supply Chain/B2B | 31 | supply, integration, chain, supply_chain, information |
| 7 | Information Privacy | 19 | privacy, information, information_privacy, research, profiling |

This demonstrates how bigrams are important in capturing core research concepts that cannot be expressed using only single words. To give more information, for each cluster, a broader set of the top 50 terms were plotted in word cloud. Word or tag cloud visualization (Kaser & Lemire, 2007) is a text analytic technique (Heimerl et al., 2014) that places terms together in a word visualization, where terms are colored and sized by relative popularity, making common terms more conspicuous. Word clouds for each of the clusters are given in Figure 2.

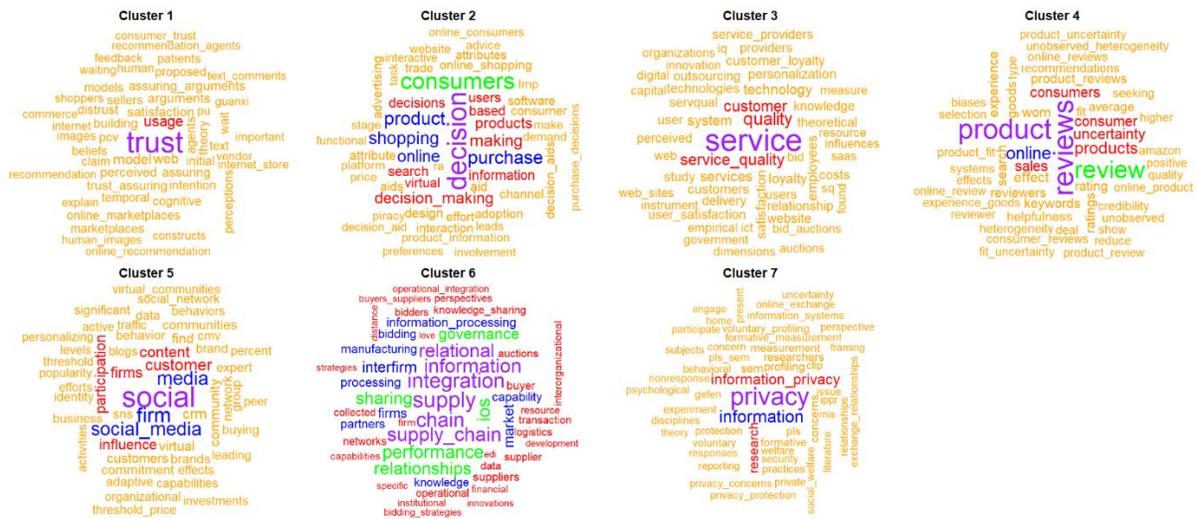

**Figure 2. Word/Tag Cloud of Clusters**

Using the information in the top 5 terms and word clouds, the clusters were named as Consumer Trust, Consumer Decision Making, Services & Loyalty, Online Reviews, Social Media and UGC (User Generated Content), Supply Chain Management/Business-to-Business, and Information Privacy. All these clusters represent distinct research areas, with some being more focused towards marketing topics and others towards IS topics. All have some focus on the "customer", whether in a marketing context (e.g., consumer decision making) or an IS context (e.g., information privacy). Most work is focused on the B2C (business-to-consumer) domain, but there is a distinct Supply Chain/B2B (business-to-business) cluster that contains terms related to supply chains, interfirm relationships, and performance. The clusters, excluding the more strongly IS focused privacy and trust clusters, show close concordance with clusters generated for a systemic review of "marketing analytics" topics in marketing journals (Iacobucci et al., 2019), giving additional face validity.

Each of the seven clusters was dealt with separately in the review. The articles selected in a cluster form the core of each "mini" literature review. In some of the larger clusters, such as consumer decision making, some of the less "core" articles were omitted in order to make the review manageable. In addition, given the asymmetry of citation patterns between marketing and information systems, leading to most of the selected journal articles being from information systems, for each article, several citing articles from the opposite discipline were selected to balance the review.

### *Selecting Papers for the Systematic Review*

For parsimony and to keep review length reasonable, a subset of around 20-25 papers was selected for each cluster (more papers were selected for the large, 94 paper cluster two). These papers were selected to give a broad representation of the work in the cluster. Pure literature reviews, methodological articles, similar work (by same author), and some work with low citation impact were removed. A summary table was created for each cluster, with a row per paper. For each paper, a short description (including journal acronym) was given under the discipline column.

## Discussion of Clusters

This section contains an analysis and discussion of the clusters developed in the previous section. To help explore the citations between disciplines, for each cluster paper, several citations for the paper from the opposite discipline were examined and then used to help form the analysis and discussions on the clusters given in the next section. In addition, cross-authorships in the clusters were analyzed and the number of cross-authorship papers between marketing and IS in each cluster was calculated.

### *Cluster One: Trust and Reliability*

All the selected papers in this cluster are from IS. Two of the twenty-two selected papers have cross-authorship between marketing and IS (Geissler et al., 2001, McKinney et al., 2002). The majority of the papers are behavioral and utilize survey data, with data analyzed through SEM.

The focus of the IS work in the cluster is on trust, with an emphasis on the issues of trust and privacy raised by the rise of the internet and e-commerce. Studies have highlighted the role of platform (website) design and cognitive factors in shaping user behavior in online platforms. Several papers examined the role of trust and risk beliefs as drivers of purchase intention in online platforms and offered different theoretical models for the conceptualization of the trust in online shopping. From the design perspective, IS work has shown that the perceived quality (Lowry et al., 2008; McKinney et al., 2002; Zahedi & Song, 2008), perceived complexity of the website (Geissler et al. 2001), presence of visual content that promote social presence (Cyr et al., 2009), and cultural factors (Sia et al., 2009) are important contributors to the formation of trust.

The cited marketing work includes foundational behavioral work that is used to help explain and conceptualize trust. Examples include work on message framing (Maheswaran & Myers-Levy, 1990), stimuli-response (Jacoby, 2002), social dynamics (Bearden & Etzel, 1982), cognitive dissonance (Cohen & Goldbery, 1970), and consumer risk (Bettman, 1973). There is also work on e-commerce adoption trust, which is predominantly from the focus of consumers (e.g., Schlosser et al., 2006; Yoon 2002) and strategy focused work on trust in buyer-seller relationships (e.g., Morgan & Hunt, 1994).

*Cluster Two: Consumer Decision Making*

This cluster is probably the broadest, most heterogeneous cluster. It contains articles where consumer decision making is the focus, but with different application areas. Three marketing papers are included in the cluster and seven out of the thirty-four selected papers include collaborations between marketing and IS authors. There are articles in this cluster that approach decision making from a rational economic utility-based choice perspective and those that approach decision making from a behavioral perspective, emphasizing psychological effects and biases. The methodology employed in the paper reflects this, with around 50% of papers utilizing economic or econometric methods and the remainder utilizing a range of methods including SEM, Bayesian statistics, and experimental ANOVA/MANOVA.

The majority of papers in this cluster are again focused on the internet and e-commerce. However, the focus is broader than in cluster one and other factors such as perceived benefit (usefulness) are considered when examining the loyalty of customers on e-commerce platforms (Kim & Son, 2009). In the IS literature, this subject has generally been studied in the context of technology acceptance and continuous usage models (e.g., Bhattacherjee, 2001; Xu et al., 2003). The articles in this cluster mostly examined the antecedents and consequences of adoption and continuous usage of e-commerce related technologies. The IS literature has mostly emphasized design elements (i.e., navigability, visual appeal, and ease of use), behavioral and cognitive factors (i.e., habits, convenience, satisfaction, and hedonic motivation), and provided information (i.e., price value, purchase mechanism, product information, and previous users' information) as important antecedents of technology adoption and usage (e.g., Duan et al., 2009; Goh et al., 2015; Kim et al., 2005; Koufaris, 2002; Kuruzovich et al., 2008; Venkatesh et al., 2012). The IS work cited a range of consumer behavior work from the marketing literature including work on consumer goal setting (Bagozzi et al., 1992), search (Ratchford & Srinivasan, 1993), variety seeking behavior (Berger, Draganska, & Simonson, 2007; McAlister, 1982), social identity (Kleine III, Kleine, & Kernan, 1993), and updating consumer judgments (Bolton & Drew, 1991a).

The marketing papers in this cluster have a slightly different focus and look at the role of personalization and recommendation systems in usage decisions of users (De Bruyn et al., 2008; Häubl & Trifts, 2000). In addition, there are several IS papers in the cluster that look at pricing effects and the consumer demand function from an economic perspective (Aloysius et al., 2013; Geng & Lee, 2013; Granados et al., 2012; Langer et al., 2012). This work draws upon marketing work on information search and price dispersion (Cachon, Terwiesch, & Xu, 2008), bundling (e.g., Jiang, Shang, Kemerer, & Liu, 2011), and the price elasticity of demand (Bijmolt, Van Heerde, & Pieters, 2005).

*Cluster Three: Services and Loyalty*

This cluster is a tightly focused cluster, with an emphasis on service provision and loyalty. The IS articles focus on online service quality provision. There is an emphasis on specific IS service areas, such as software as a service and e-government services. There are two marketing articles, both in JAMS, which utilize work on technology acceptance/IS resources in a service context. Three of the twenty-two selected articles have cross authorship between IS and marketing. In terms of methodology, the majority of the work in this cluster is survey based behavioral work, utilizing scale development, analyzed with SEM. There is a significant minority of econometric work.

While the focus in cluster two is on examining the role of satisfaction in e-commerce, the papers in the third cluster focus on the quality of service as an antecedent of satisfaction. Although most of the studies in this cluster are IS papers, they rely strongly on the service marketing literature and the "SERVQUAL" service quality framework (e.g., Parasuraman, Zeithaml, & Berry,1988; Pitt et al. 1995) used to measure various aspects of service quality. Studies have shown that regardless of how the service is delivered to consumers, either as a support feature (Cenfetelli et al., 2008; Susarla et al., 2003) or as a main platform product (Tallon, 2010; Tan et al., 2013; Xu et al., 2013, 2014), its quality can be decisive in the satisfaction, and ultimately in the retention of customers. Studies have also highlighted the role of human services that supplement and sustain electronic services (Akçura & Ozdemi 2017; Ba et al. 2010), organizational factors, such as employee satisfaction (Hsieh et al., 2012), and service personalization (Xu et al., 2014).

While the focus of the cited marketing work is on SERVQUAL and related measures of service quality, a range of other work is utilized, including behavioral work on consumer expectations (Oliver, 1980) and consumer decision processes (Cox & Rich, 1964), and strategy work on aspects of customer centricity (e.g., Shah, Rust, Parasuraman, Staelin, & Day, 2006), service technology use (Bitner, Brown, & Meuter, 2000), and service failure (Holloway & Beatty, 2003).

*Cluster Four: Online Reviews*

Since the first online review and reputation systems were created by e-Commerce retailers such as Amazon.com in the late 1990s, online reviews have been of great interest to both marketing and IS researchers. Online reviews sit at the boundary of several important topics, including eWOM (electronic word of mouth), reputation systems and management, user-generated content, and social media. Four of the twenty-four selected articles have cross authorship between IS and marketing.

The papers in this cluster focus mainly on the actual reviews, as well as review and reviewer characteristics. The older work in the cluster utilizes review characteristics, such as volume and valence, but some of the newer papers include increasingly sophisticated text mining work. Over 75% of the chosen articles utilize econometrics (usually panel data econometrics), and the remainder include a few behavioral survey research/SEM articles and experimental work analyzed with ANOVA/MANOVA. This mix is unsurprising given that the focus of this cluster is predominantly on empirical analysis of reviews that are collected over time. Apart from two marketing papers, all the papers in this cluster are from IS journals.

Cluster two contains IS work that examines how perceived usefulness was one of the key factors in the loyalty and continuous usage of online platform users. Several of the IS studies in this cluster (cluster four) link to this work by arguing that product reviews posted by other customers on online platforms could significantly boost the perceived usefulness of online shopping platforms (Benlian et al., 2012; Kumar & Benbasat, 2006; Kwark et al., 2014). Beyond the general emphasis of the papers in this cluster on perceived usefulness, the topics of the articles in this cluster can be classified into four categories: 1. Combination of consumer reviews and marketing campaigns to improve sales (Li, 2016; Lu et al., 2013; Shen et al., 2015), 2. The effect of reviews on market competition (Kwark et al., 2014; Li, 2017; Li et al., 2011), 3. Analysis of review features that make them effective and credible (Cheung et al., 2012; Huang et al., 2013; Jensen et al., 2013; Li et al., 2017; Ma et al., 2013; Mudambi & Schuff, 2010; Zhou & Duan, 2016), and 4. Factors that influence the process of creating customer reviews (Dellarocas et al., 2010; Hong et al., 2016; Ho et al., 2017).

The work listed above utilizes a range of behavioral work from marketing to help explain consumer reviewing behavior and consumer response to reviews. This work includes work on attitude formation (Rucker, Tormala, Petty, & Briñol, 2014), consumer satisfaction confirmation/disconfirmation (Bolton & Drew, 1991b), cognitive fit theory (Huang & Chen, 2006), consumer processing (Petty, Cacioppo, & Schumann, 1983), two-sided versus one-sided appeals (Crowley & Hoyer, 1994), and choice processing (Scheibehenne et al., 2010). There is also a good deal of cited work in marketing on WOM (word of mouth)/eWOM. This may indicate differing emphases between IS and marketing. While IS research has focused on the concepts of online reviews and review systems, marketing places online reviews within the long existing WOM research stream (e.g., Arndt, 1967) and considers online reviews to be a type of eWOM. Key marketing work cited on the subject of WOM includes work on favorable/unfavorable reviews (Chevalier & Mayzli, 2006), work relating WOM to marketing strategy (Chen & Xie, 2005), work on WOM source credibility (Brown et al., 2007), and work on social motivations for eWOM (Sridhar & Srinivasan, 2012). There is also a quantitative stream of research on online review dynamics, which includes a cluster paper (Godes & Silva, 2012) and several cited papers (Moe & Schwidel, 2012; Moe & Trusov, 2011).

*Cluster Five: Social Media and UGC (User Generated Content)*

This cluster has papers that focus on social influence and users' behavior in online social environments. The work in this cluster touches on online reviews and eWOM, but gives a broader view of the online social infrastructure and includes work channels, such as online social networks, blogs, microblogs, and virtual communities, which can all play a role in the dissemination of product and service information, as well as user experiences.

Out of the twenty-five selected papers in the cluster, two are in marketing and there are three papers with collaboration between IS and marketing researchers. Around 60% of the papers use econometrics, but there are a fair number of behavioral survey research/SEM articles.

Many of the IS studies focus on relating social participation behavior to purchase behavior (Geva et al., 2017; Kuem et al., 2017; Lee et al., 2015; Oestreicher-Singer & Zalmanson, 2013; Rishika et al., 2013; Xie & Lee, 2015) and brand perceptions (Luo et al., 2013). There is work on comparing the relative effects of social content and marketer-generated content (Goh et al., 2013) on brand sentiment. There is also work highlighting the role of social influence in emerging areas of e-commerce, such as group-buying and music communities (Dewan et al., 2017; Kuan et al., 2014). The other major set of studies in this domain examines the factors that impact the degree of contribution of users to social media and other virtual communities (Chen et al., 2014; Ray et al., 2014; Tsai & Bagozzi, 2014).

This cluster contains several social network modeling papers. Zhang et al. (2016) utilized community detection and sentiment analysis to build brand relationship networks and recommend brands to users. The two marketing papers in the cluster are both mathematical modeling papers that utilize social data to build a predictive model for consumption (Chung et al., 2016) and a network effects model of content popularity (Ransbotham et al., 2012).

The cited marketing work includes both behavioral and mathematical modeling work on the dynamics of social influence within social networks and online communities (e.g., Ansari et al., 2011; Bagozzi & Dholakia, 2002; Brown, Broderick, & Lee, 2007; Godes & Mayzlin, 2004; Trusov at al., 2010), which has been used by IS researchers to build theories on user interaction with social media platforms. There is a range of basic behavioral work cited on topics such as identity motivations (Oyserman, 2009), communications valence (Mizerski, 1982), and the relationship between consumer emotions and behavior (Soscia, 2007). In addition, when relating social behavior to branding and purchasing, a wide variety of marketing strategy material is cited by IS researchers, including work on consumer equity (Rust et al., 2004), brand involvement (Beatty & Kahle, 1998), and consumer investment/switching costs (Burnham et al., 2003), along with work that relates customer satisfaction (e.g., Anderson et al., 2004) and firm performance (Luo, 2007) to brand value.

*Cluster Six: Supply Chain Management/Business-to-Business*

This cluster is quite separate from the other clusters and includes a range of business-to-business relationship work at the interface of IS, marketing, and supply chain management. Three of the eighteen selected articles have both marketing and IS authors. The cluster contains both behavioral research and econometrics research in approximately equal proportions.

There are four marketing articles, all from JAMS. These articles look at IS/marketing integration (Davis-Sramek et al., 2010; Nakata et al., 2011), IS competency (Davis & Golicic, 2010), and information exchange (e.g., Kim et al., 2006), all from a marketing channels perspective.

In the IS articles, there is an emphasis on the role of information system infrastructure in the development of inter- and intra-company relations. There is a particular focus on the role of interorganizational information systems (IOS) (Chatterjee & Ravichandran, 2013) and EDI (electronic data interchange) (Son et al., 2005) in enabling relationships between channel partners. Articles cover topics such as information sharing (Dong et al., 2017; Grover & Saeed, 2007; Saraf et al., 2007), information processing (Premkumar et al., 2005; Wang et al., 2013) and flow of information (Klein & Rai, 2009; Patnayakuni et al., 2006) between organizations.

The IS articles listed above utilize key work in marketing on B2B relationships and marketing channels. Examples include work on building commitment in B2B channels (Anderson & Weitz, 1992), influence strategies in buyer-seller relationships (Fraziee & Rody, 1991), B2B relationship governance (Ghosh & John, 1990; Heide, 1994), and relationship orientation, including strategic orientation (Matsuno et al., 2002; Zhou et al., 2005) and long/short term orientation (Ganesan, 1994).

*Cluster Seven: Information Privacy*

Work in this cluster focuses on privacy concerns/privacy protection in e-commerce, the effects of customer profiling on privacy, and consumer responses to privacy issues.

The cluster is a strongly IS focused. There is a single marketing article in JCR (John et al., 2011), which utilizes IS work on information privacy to examine the antecedents of getting survey respondents to admit to risky or socially undesirable behavior. This article is the only one out of the seventeen in the cluster that has cross authorship between marketing and IS. The majority of the articles in the cluster utilize survey research and SEM, though there are several econometric and game theory articles.

Information privacy is a core area of IS research. The articles selected for this cluster and that cite the marketing literature focus primarily on issues involving information privacy and the consumer. Topics include the trade-off in e-commerce between using customer information for personalization/customization and privacy (Chellappa and Shivendu, 2007; Lee et al., 2011), how privacy concerns can affect consumer purchase intent (Van Slyke et al., 2006), and how a firm's recovery from a security breach affects consumer behavior (Choi et al., 2006). There is work that examines how e-commerce privacy features and certificates improve customer willingness to pay higher price premiums (Mai et al., 2010) and provide private information (Hui et al., 2007). These e-commerce "consumer privacy" articles cite an eclectic mix of articles from the marketing literature, including work on consumer incentives (Alba et al., 1997), trust in buyer/seller relationships (Doney & Cannon, 1997), service recovery (Hoffman & Kelley, 2000), price premiums (Rao & Bergen, 1992), and relating privacy to consumer trust (Phelps et al., 2001).

## Discussion and Conclusions

*The Bibliographic Review Process*

This paper set out to analyze research at the boundary of marketing and IS under the heading of the question, "who owns the data?" This was achieved by selecting marketing papers that cited a certain number of IS papers and vice-versa and then using document clustering to bring "order" to the results and to find certain "areas of collaboration" between the disciplines. Seven different research "areas of collaboration" were identified. The papers were further filtered and analyzed. Methodological, pure review articles, and "odd" articles non-congruent to the clusters were removed. Some articles with the same authorship groups and similar topics were "pruned" to make the size of the clusters manageable, while keeping a good breadth of topics.

To further examine the collaboration between disciplines, for each IS paper in a cluster, several citing marketing papers were chosen and vice-versa for marketing papers. This analysis of citations was used to help write the cluster descriptions.

*Asymmetry between Marketing and IS*

As previously noted, the cross citation of references between marketing and information systems is very asymmetric. There may be several reasons for this. First, marketing is an older discipline than IS. There is a lot of core work from the sixties and seventies that is cited by both marketing and IS. Second, marketing is a larger discipline than IS. A recent AACSB salary survey (AACSB, 2019) found approximately 50% more marketing faculty than IS faculty. Assuming equal research productivity across disciplines, this will lead to more citable research in marketing. Third, a greater proportion of academic marketing research is positioned closer to being core social science research than for academic information systems research. This is often to the chagrin of more applied strategy and practitioner focused researchers in marketing (e.g., Baines et al., 2009; Piercy, 2002; Tapp, 2004). However, this orientation has created a large repository of ``basic'' research in consumer behavior in areas such as information processing, consumer search, and social persuasion, which is cited in a similar manner to psychology or sociology articles and can be applied to a range of IS research scenarios, including IS user adoption behavior, interface design, and relationship management. In fact, the area of marketing consumer behavior research has been the "invisible man" of this review, in that virtually no marketing consumer behavior papers were found at the boundary of marketing and IS, but many of the IS papers cite work from marketing consumer behavior.

In addition, given the importance of methodological correctness in academic research, most papers cite a range of methodological papers to give justification for the methods used in the paper. These articles are from both inside and outside of the business disciplines. Again, there is a degree of asymmetry, with more of these articles from marketing, particularly in the areas of structural equation modeling (SEM), choice modeling, segmentation, and scale development.

While the citation patterns differ between disciplines, the underlying methodologies employed in the selected articles are similar. Most papers utilize behavioral experiments, survey research with SEM models, econometric models on real world data, analytical game theory models, and other (statistical/optimization) quantitative methods. In both disciplines, there has been a slow move from survey research analyzed with SEM to econometric estimation and experimental work, though this trend has been faster in marketing than in IS.

*Areas of Collaboration*

The original intent of this study was to examine collaboration over time between marketing and IS. However, as per Figure 1 and the associated discussion, very few papers were selected before the year 2000. The growth of collaboration between marketing and information systems is directly tied to the growth of e-commerce and the internet. Most of the topics discussed in the paper relate to some aspect of e-commerce or associated topics. Six out of the seven clusters focus in some way on the consumer. In fact, the locus of collaboration in most of the clusters is on consumer behavior. Given the asymmetry in citations, the papers in the article clusters are overwhelmingly from IS. Some of the clusters, for example the "Online Reviews" cluster and the "Social Media and UGC" cluster, form distinct areas of research that are of interest to both marketing and IS and these clusters have significant collaboration and cross-authorship between disciplines. Other clusters, for example the "Trust and Reliability" and "Privacy" clusters form subsets of larger IS areas of research, but have a specific focus on the customer and engage marketing by citing work on consumer behavior and the consumer shopping environment. The "Consumer Decision Making" cluster is the largest cluster in terms of papers. The papers in this cluster contain an eclectic mix of topics, but there is a strong core focus on how consumers make decisions in online environments and the predominantly IS papers heavily cite marketing consumer behavior papers.

The "Services & Loyalty" cluster arose due to a specific business need. In corporate information systems, with the growth of the internet, there has been a move away from a model where computer software is considered a product and towards one where software is distributed as service and data are stored remotely in the internet "cloud". The topic area of "software as a service" has become of an area of interest for IS researchers (e.g., Benlian & Hess, 2011), along with the wider infrastructure of new "Web 2.0" technologies, which include "software as a service", along with social applications, blogs (e.g., Chau & Xu, 2012), and other collaboration tools (Andriole, 2010). IS work in this area has found and adapted existing frameworks in services marketing for analyzing and measuring services. There is particularly heavy use made of the SERVQUAL framework (Parasuraman et al., 1988) and associated measures.

The "Supply Chain Management/Business-to-Business" cluster differs from the other clusters in that the focus is on B2B work at the intersection of marketing, IS, and supply chain management. Again, this cluster is driven by changes in business practice. Supply chains and B2B channels have seen a significant move to technology-mediated environments and increased automation, with B2B e-commerce channels and technologies such as EDI and IOS. These technologies are of interest to IS researchers and given that these technologies are utilized in an existing B2B/supply chain environment, IS work cites marketing and supply chain work that examines and explains this environment.

Overall, research at the boundary of marketing and IS has been driven by e-commerce technologies that are of interest to both IS and marketing researchers and give rise to cross-disciplinary research questions at the intersection of the consumer and technology. In addition, the move of software to a service environment and the increased use of technology and e-commerce in B2B have further driven cross-citation between marketing and IS. These issues are discussed further in the theoretical framework given in the next section.

*Theoretical Framework for Collaboration*

This review examines specific areas of collaboration between marketing and IS and gives prescriptive ideas for moving research at the boundary of IS/marketing forward, both from a practitioner perspective and an academic research perspective. A major aim of this review is to "shine a light" on these areas of collaboration and make this knowledge available to both IS and marketing researchers. These areas of collaboration are part of a much broader academic research infrastructure that incorporates both the social and behavioral and the physical and natural sciences. Carley et al. (2017) visualized this network infrastructure, showing, for example, that quantitative business disciplines are intermediate to the social and behavioral sciences and the mathematics portion of the hard sciences.

The previous section outlined the areas or loci of collaboration between marketing and IS and noted the centrality of e-commerce topics in generating research at the boundary of IS and marketing. Most of the areas of collaboration occur in the sphere of e-commerce and newer "Web 2.0" technologies. Unsurprisingly, much of this work is motivated by the interaction between consumers and/or businesses with these new technologies. Some research topics are broad based IS research topics. For example, IS researchers (Bélanger & Crossler, 2011) have noted that information privacy is a "key construct" in IS research. The area of collaboration found by this review is a subset of this area that engages with consumer-based topics and includes consumer behavior in its theoretical models. Conversely, the "online reviews" area is an area with a much stronger overlap between marketing and IS, with cross-authorships between the disciplines and with researchers in both disciplines concerned with both the technology aspects of online reviews and the consumer behavior aspects.

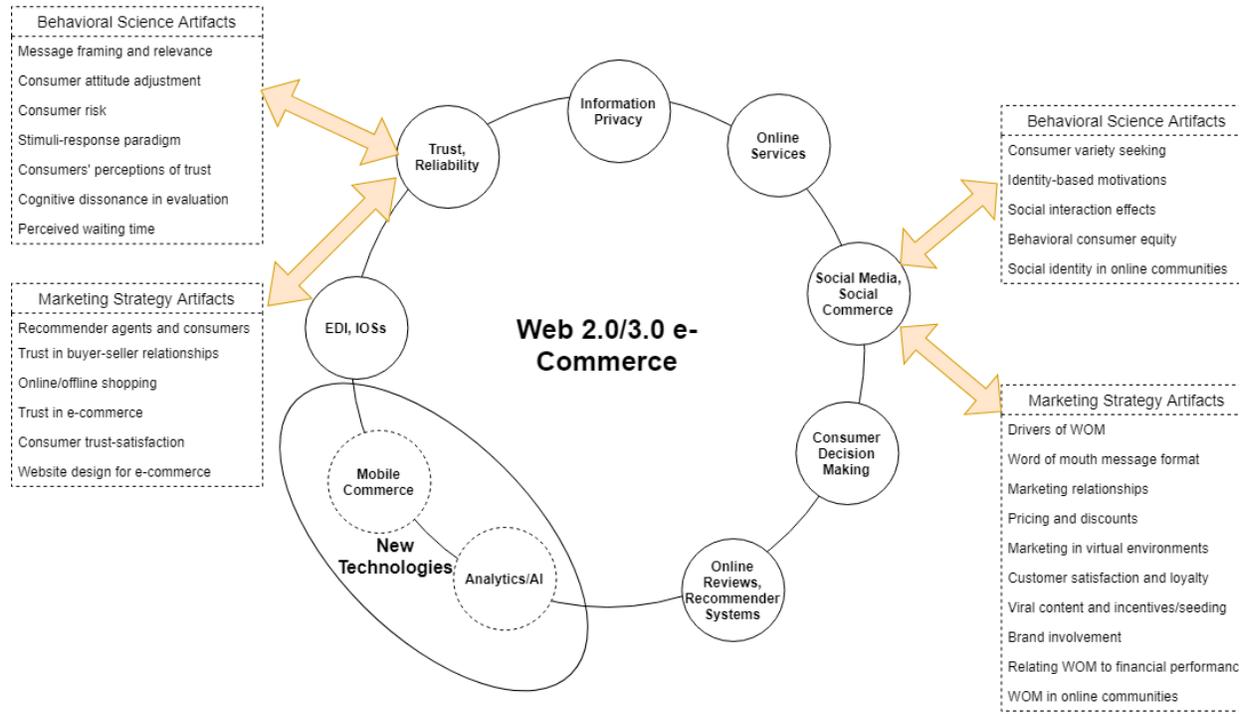

**Figure 3. Conceptual Model of Web 2.0/Web 3.0 Generated Research**

As new technologies are developed, particularly in the realm of e-commerce, then these technologies attract the interest of both marketing and IS researchers. A new e-commerce technology may have aspects that pertain to core areas of IS research such as system design and implementation, technology acceptance (Lee et al., 2003), knowledge management (Alavi & Leidner, 2001), and human-computer interaction (HCI) (Zhang & Li, 2004). Likewise, marketing researchers may be interested in marketing aspects of the new technology, such as how the new technology impacts customer service (e.g., Meuter et al., 2000), branding, and online social interactions (e.g., Berthon et al., 2012) or how it can be leveraged gain competitive advantage by increasing customer value (Parasuraman, 1997). However, given the context of e-commerce technology, there are strong links between the IS and marketing aspects of the technology and these links form the nexus of collaboration between IS and marketing. For example, in the context of social web technologies, technology user acceptance and the HCI aspects of system design are strongly related to the preferences of the users, who are the "consumers" in a marketing context. This view of new technology driving emerging topics at the boundary of IS and marketing is validated by a longitudinal bibliographic analysis of digital marketing topics in Krishen et al. (2021), who showed how academic topics in e-commerce and internet marketing have been developed in conjunction with advances in e-commerce technology, starting with topics such as customer trust and internet marketing in the late 1990s and going through to topics such as virtual reality and the shared economy in 2019.

From a knowledge management perspective, new technologies provide opportunities to acquire, create and harness new knowledge to support decision making (McCall et al., 2008), while from a data-based marketing perspective (e.g., Wedel & Kannan, 2016), data are used in conjunction with analytic methods to improve marketing decisions on product, price, place, and promotion. The definition of "knowledge" in knowledge management is a little broader than pure "data", though data analytics can be brought into a knowledge management framework; for example, Pauleen and Wang (2017) built a conceptual model where data analytic techniques are utilized to create new knowledge that can be used to aid decision makers and gain insight into business processes. Likewise, Khan and Vorley (2017) noted that text analytics can help structure, visualize, and analyze data, which can improve the knowledge management process. Given the rapid increase in the size and availability of data, in both IS (e.g., Abbasi et al., 2016) and marketing (e.g., Erevelles et al. 2016), there is a realization that "big" and unstructured data, which can be characterized by the volume, variety, velocity, and veracity of the data (e.g., Miele & Shockley, 2012), present unique challenges and data-analytic techniques and processes must be adapted to deal with these challenges.

Figure 3 contains a simplified conceptual model of the areas of collaboration between IS and marketing. At the heart of this model are the new areas of research enabled by new e-commerce (and related) technologies. In the real world, each area of IS/technology research has an interface to marketing through the consumer (or business) and some aspect of marketing (e.g., product, price, place, promotion). Work at this interface can utilize research artifacts from behavioral research and from marketing strategy. In Figure 3, examples of these artifacts are given for trust and reliability (summarizing Table 7) and social media/UGC (summarizing Table 10). The artifacts for the remaining topics are left for the sake of brevity. Several emerging topics are included to illustrate how topics related to new technologies can be included.

Consider an example of how this framework could be used to drive future research. Brain/computer interface technology is an emerging technology, which could profoundly alter how humans interact with our environment. Applications already exist in the medical arena, for dealing with rehabilitation and brain injuries and there are a range of commercial applications including and there are a range of commercial applications including brain training or "Fitbit for the brain" and the ability to seamlessly connect to a range of systems and data sources (Zhavoronkov, 2021). This technology raises a whole host of privacy issues, given the automatic "unconscious" access that systems will get to the human brain. Work dealing with privacy and on building consumer trust and trust-assuring arguments (clusters one and seven) could be applied to the specifics of this new technology and will rely on theoretical artifacts from marketing in the previously noted areas of message framing, stimuli-response, social dynamics, cognitive dissonance, and consumer risk. All new technologies involve consumer decision making (cluster two). For any new technology, there are consumer decisions on adoption and behavioral artifacts from marketing in areas such as information search, variety seeking, and goal setting would be key to explaining decision making in an IS context. This technology could strongly affect the e-commerce realm. Brain/computer interface technology could "shortcut" some of the functionality provided by the internet (Berger, 2021) and thus affect how consumers utilize many of the features of e-commerce systems and how they provide user-generated content and reviews (clusters four and five). Work on social networks and consumer eWOM would be given new context in an environment where communication was more direct and theories on social processes from marketing and the behavioral sciences could be utilized. Finally, there may be a service aspect to the implementation of the new technology, where consumers subscribe to different data sources/connected applications in a similar manner to software as a service. Service-based work could draw on software as a service work and more general service topics from marketing such as service quality measurement (e.g., SERVQUAL) and service recovery.

*Implications and Recommendations for Practice*

The major focus of this paper has been on the increasing overlap between IS and marketing in academic research. A similar phenomenon exists in practice. Today's marketers need to have a strong understanding of technology in order to operate in an e-commerce environment. A HBR magazine article (Brinker & McLellan, 2014) noted the rise of executive level marketing technologists, who work at the boundary of marketing and IS and who understand how to best deploy technology to align with marketing needs. Any analysis of current marketing jobs will find a whole host of jobs at the boundary of marketing and IS, including, but not limited to, social media manager, digital marketing manager, search engine optimization (SEO) manager, online community manager, A/B tester, UX (user experience) designer, digital marketing strategist, and a range of research/analytics positions that require data and technology skills. In addition, many traditional marketing jobs, require technical knowledge and skills. For example, technology has changed the job of a media planner. Around 50% of media are now consumed in digital form and in the key 18-34 demographic this percentage is much higher (Nielsen, 2020). Media planners now need to understand and account for different digital and online channels and consumption patterns. The trend towards "hybrid" technical marketing jobs is predicted to continue. A report by Cognizant (2019) identified twenty-one emerging, mostly technical, marketing jobs, including jobs that utilize current technologies, such as "data ethnographer" and "loyalty engineer", to jobs that are more speculative, such as "head of bot creative".

The marketing technology environment described above is bound to generate problems that are relevant to both marketing and IS researchers and can utilize knowledge from both areas. Disquiet from marketing academic researchers regarding the current relevance of academic research was noted in previous sections. Similar self-reflection has been carried out by IS researchers (e.g., Hassan, 2014; Kock et al., 2002; Moeini et al., 2019; Rosemann & Vessey, 2008) and several researchers have emphasized the need for more programmatic and design science research based on academic and industry collaboration (Gregor & Hevner, 2013; Hevner et al. 2004; Ram & Goes, 2021). Practical, industry-driven problems in the emerging marketing technology environment could provide impetus both for improved relevance and for collaboration between marketing and IS researchers. There have been several initiatives to help do this, including the Wharton Customer Analytics Initiative (WCAI, 2020), in which businesses provide datasets and pose questions for academic researchers and the INFORMS-Marketing Science Institute Practice Prize (Lilien,

Roberts, & Shankar, 2013), which rewards academic-practitioner collaborations with recognition and publication in the Marketing Science journal. Such initiatives, along with journals that bring academic insight to practitioners, such the general Harvard Business Review and Sloan Management Review journals, and discipline specific outlets such as MISQ Executive and INFORMS Interfaces, combined with incentives for faculty to publish in these outlets (Lilien, 2011), and collaborative IS-marketing research teams that mirror the collaboration in industry, could all help bridge the academic-practitioner divide in e-commerce and marketing technology. From an industry perspective, there is an increasing tradition of informal meetup groups in the technology and analytics space that help participants network, learn new skills, and even engage in projects to help non-profits (France & Ghose, 2019). If utilized by academic researchers, such groups could prove valuable to future collaboration.

*Implications and Recommendations for Academia*

The conceptual model in Figure 3 illustrates the areas for collaboration between marketing and IS, primarily driven by new technological advances. Of particular interest is the area of e-commerce. Over the last 20 years, e-commerce technologies have been a driver of research at the boundary of marketing and IS. In an editorial for MIS Quarterly, Straub (2012) noted that "With the advent of very viable Web-based enterprises from Amazon to Google to e-Bay, we have seen a much tighter relationship between marketing constructs/theories and IS" and notes there is potential for increased collaboration and increased cross-citations between the marketing and IS disciplines. While there has been a growth in cross-disciplinary collaborations and cross-citations, there is a strong asymmetry in citations, with many more citations flowing from marketing to IS than in the opposite directions.

The asymmetry in citations between marketing and IS is noted in this review as a state-of-being or ontology. Given the previously described issues of relative discipline size, age, and the larger number of methodology/purer behavioral articles in marketing, it may be impossible to achieve complete symmetry between the disciplines, but several methods could help redress the balance.

Literature reviews with authors from both IS and marketing in areas of mutual interest could help expose researchers to relevant work from the other discipline. Several recent reviews have done this. For example, an IS/marketing cross-authored paper by Dwivedi et al. (2020) reviewed digital marketing from both IS and marketing perspectives. Similar collaboration could be extended to conferences and workshops, as suggested by Straub (2012). In addition. some authors have noted the narrowness of business doctoral programs (El Gayar, 2006; Roach et al., 1994). While in doctoral programs, students should be encouraged to take seminars in non-core, related areas in order to expose them to theory in those areas.

For each cluster, a set of artifacts used by one discipline (usually IS) from a second discipline (usually marketing) was identified and described. However, given the broad nature of the review, it was not possible to make this list comprehensive. Follow-up studies could focus more narrowly on the different areas of the review and produce detailed listings of theoretical research artifacts that lie at the boundary of marketing and IS, perform a meta-analytic analysis of how different research artifacts validate and contradict one another, and examine the magnitude of different effects in order to create empirical generalizations (e.g., Hanssens, 2018). Such meta-analyses could be used to answer several questions. How do theories in marketing and IS overlap? What are the commonalities? Are there studies with common main effects but different moderators and mediators? Are theoretical similarities clouded by differences in language and terminology between the disciplines? This review has found some asymmetry in how theoretical artifacts from one discipline are used in the other discipline. Can more "micro" studies shine further light into this phenomenon and categorize theoretical artifacts in IS that will be useful for marketing research?

While the current review is focused on IS and marketing, other research areas may be relevant, and a technological focused review could look at boundaries with other disciplines. For example, the technology acceptance literature in IS overlaps with management and marketing channels and supply chain research overlaps both marketing and operations. In addition, theories from other disciplines, such as management, have been independently incorporated into marketing and IS. For example, the resource-based theory of the firm from management has been utilized in both information systems (e.g., Kearns & Lederer, 2003; Wade & Hulland, 2004) and in marketing (e.g., Capron & Hulland, 1999; Fahy & Smithee, 1999). A more detailed conceptual model could incorporate some of these more complex "second-level" associations.

Finally, the "Theoretical Framework for Collaboration" gives a model for technology and e-commerce research collaboration. Currently, this model is purely conceptual. Further empirical research could help ground and refine this model with data by taking a "technology focused view" of article publishing and citation patterns for specific new e-commerce technologies and examine how theory is built across disciplines for each technology.

# Appendix A: Cluster Papers

**Table 6. Cluster One: Trust and Reliability**

| Journal | Description |
|---|---|
| ISR | McKinney et al. (2002) built a model where web customer satisfaction is a function of service quality (adapted SERVQUAL) and information quality (understandability, reliability (including trustworthy) and usefulness) disconfirmation. |
| ISR | Kim and Benbasat (2006) ran a behavioral experiment, showing that trust-assuring statements from e-commerce providers are best augmented by data and statements that explain these data. |
| ISR | Pavlou and Gefen (2005) built a model of the antecedents (e.g., past buyer experience and seller performance) and consequences (e.g., trust, perceived risk, buying intentions) of psychological contract violation in e-commerce. |
| ISR | Pavlou and Dimoka (2006) examined how in online marketplaces (e.g., eBay) the credibility and benevolence of seller feedback comments affect trust in the sellers and thus affect the price premiums that sellers can charge. |
| ISR | Kim et al. (2009) showed that increased trust in an e-commerce provider lowered perceptions of risk, increased perceived benefits, increased willingness to purchase products, and increased satisfaction and loyalty. |
| ISR | Liu and Goodhue (2012) found that consumers have a trust threshold, and once that threshold is passed, no longer consider trust as a factor when considering purchases or repurchases. |
| JMIS | Pennington et al. (2003) built a model of how indicators of system trust (seals, guarantees, and ratings), along with vendor reputation, affect consumer trust, attitude, and purchase intent for a vendor. |
| JMIS | Wang and Benbasat (2008) examined the process of building consumer trust in recommendation agents (RAs) and modeled how different aspects of trust affected overall trust in RAs. |
| JMIS | Zahedi and Song (2008) built both static and dynamic models of trust in healthcare intermediaries. In the dynamic model, satisfaction updates prior beliefs of ability, benevolence, and integrity. |
| JMIS | Lowry, Vance, Moody, Beckman, and Read (2008) explored how less well-known e-commerce sites can use a combination of high-quality websites and co-branding with well known partners to increase customer trust. |
| JMIS | Charki and Josserand (2008) built a model that examined how online reverse auctions can lead to "under-socialization" of buyer-seller relationships, which combined with unethical behavior, a lack of transparency, and technical issues can lead to a lack of trust. |
| JMIS | Kim and Benbasat (2009) examined the interaction between trust-assuring arguments and product pricing for e-commerce. For low prices, a third-party claim resulted more trust than a store claim (converse for high prices). |
| JAIS | Geissler et al. (2001) showed that there is an inverse-u shaped relationship between website complexity and website effectiveness, with moderately complex websites being the most effective. |
| JAIS | Wang and Benbasat (2005) built a model of how trust in RAs (competence, benevolence, and integrity) interacting with the perceived usefulness and ease of use of the agent impact consumers' intentions to adopt the agent. |
| JAIS | Son et al. (2006) examined how a range of factors including uncertainty (and risk), along with asset specificity, efficiency, easy of use, and effectiveness affect intention to use infomediaries/seller RAs. |
| JAIS | Sun et al. (2013) created a model of banner processing, where the salience created by structural factors (e.g., animation, color, & density) and semantic factors (e.g., entertainment & credibility) help inform aspects of banner attention (e.g., recall, arousal, eye-tracking). |

| Journal | Description |
|---|---|
| MISQ | Komiak and Benbasat (2006) built a model of RA adoption, where perceived personalization and familiarity with an RA leads to cognitive trust in its competency and integrity, which leads to emotional trust and adoption. |
| MISQ | Cyr et al.(2009) showed how incorporating human images with facial expressions into websites increases consumer-rated image appeal and perceived social presence, which leads to more trust in the website. |
| MISQ | Sia et al. (2009), in a cross-cultural context, examined how peer consumer recommendation and portal affiliation of a website affect consumers' trust and thus attitude and buying intention and behavior. |
| MISQ: | Lee et al. (2012) showed how the use of a filler interface with text and image components while a user waits for an e-commerce task to complete (e.g., a ticket purchase) affects a user's cognitive absorption, perceived wait time, website appraisal, and use intentions. |
| MISQ | Ou et al. (2014) built a model of how in China effective use of engagement tools (instant messenger, message box, feedback system) can increase customer perceptions of interactivity and presence, which leads to "Swift Guanzi", trust, and repurchases. |
| MISQ | Fang et al. (2014) examined how the perceived effectiveness of e-commerce governance and institutional mechanisms moderates the relationships between vendor satisfaction and vendor trust (+) and the relationship between vendor trust and repurchase intent (-). |

**Table 7. Cluster Two: Consumer Decision Making**

| Journal | Description |
|---|---|
| ISR | Koufaris (2002) combined website design (ease of use, search behavior) and consumer behavior (e.g., perceived control, shopping enjoyment) constructs to build an integrated model of online purchase and return behavior. |
| ISR | Kim et al. (2005) built a model where a consumer's past use and value judgments (utilitarian, hedonic, and social) for a website impact usage intention and found that past use makes consumer decision making less evaluative. |
| ISR | Hong and Tam (2006) examined how consumer psychographics (perceived enjoyment, need for uniqueness) and demographics, combined with social influence and usage characteristics affect the attitude and adoption of online services. |
| ISR | Kuruzovich et al. (2008) built a model of how the price and product information provided by online information sources interact with consumer needs for this information to influence online consumer behavior. |
| ISR | Parboteeah et al. (2009) explored how task relevant cues (e.g., navigation aids) and mood relevant cues (e.g., appealing graphics) in websites combine to affect consumers' perceptions and impulse purchases. |
| ISR | Granados et al. (2012) found that when consumers book travel through online channels there is higher price elasticity of demand. This effect is higher for opaque travel agencies (e.g., Hotwire). |
| ISR | Wattal et al. (2012), utilizing a database of promotion emails, found that product-based personalization had a positive effect on response, but a personalized greeting had a negative effect (moderated by familiarity). |
| ISR | Langer et al. (2012) examined how e-channel purchase behavior and the interplay of channel inertia, price sensitivity, and diversity of product assortment account for buyer preference heterogeneity. |
| ISR | Aloysius et al. (2013) built a game-theoretic model of sequential pricing, where, given two linked products A and B, the price for B is set once a purchase decision is made for A. This strategy can increase profit over a simple/mixed bundling strategy. |
| ISR | Ding et al. (2015) utilized a range of e-commerce marketing and consumer stimuli (e.g., past user behavior, browsing activities) to gauge purchase intent, which is then used to adapt the webpages shown to a consumer. |
| ISR | Baird et al. (2016), in the context of software sales, examined how altering pricing, extending the product range (up or down), and introducing piracy constraints affected consumer willingness to pay for high and low valued software. |
| ISR | Bhatnagar et al. (2017) utilized a multi-period, multi-state hazard model to examine how the advertising method a consumer uses to reach a website affects the time spent on the website, which affects the duration of future visits and purchase probability. |
| ISR | Tan et al. (2017) used movie rental data to show that as variety increases, consumer demand becomes more concentrated, with more demand for top products and less for niche products. |
| JMIS | Thong and Yap (1998) utilized Hunt and Vitell's ethical decision-making theory to examine the interplay of deontological and teleological processes employed by IT professionals when deciding whether to illegally copy software. |
| JMIS | Kohli et al. (2004) built a model showing how intelligence, design, and choice decision-stage constructs of the buying process imply cost savings, time savings, and satisfaction for e-commerce users. |
| JMIS | Sen et al. (2006) found favored vendor price, perceived price dispersion, and awareness of online search tools were significant predictors of consumers' online search strategy (no search, shopping agent, search engine, mixed). |
| JMIS | Köhler et al. Dellaert (2011) found that e-commerce decision aids offering concrete (abstract) advice were most likely to persuade consumers when the consumer decision was immediate (distant). |
| JMIS | Geng and Lee (2013) built an analytical model of consumer search and pricing strategy, where consumers can choose from a legitimate or pirated channel. Consumers can be shoppers or non-shoppers (with higher search costs) and search sequentially in-channel. |
| JMIS | Lee and Tan (2013) built Bass style diffusion models (with innovators/imitators) for diffusion of freeware and trialware software and looked at how the diffusion process is affected by third party/user ratings. |
| JMIS | YI et al. (2015) found that when users were given product presentations with no, restricted, and full interaction, interaction improved consumer enticement and for high knowledge consumers, enticement was highest for restricted interaction. |
| JMIS | Moody and Galletta (2015) built a model of how information scent (i.e., consumer cues), along with consumer time constraints, affect stress and consumer website performance and satisfaction. |
| JMIS | Berger et al. (2015) ran a conjoint analysis to examine how format affects (WTP) willingness to pay for online news subscriptions and found that a low WTP for online content was due to a preference for traditional paper formats. |
| JMIS | Clemons et al. (2016) ran a cross-cultural study of consumer trust and WTP in online retailing in the US, Germany, China, looking at the interaction between country and the level of quality insurance (none, quality promises, quality promises backed by third party). |
| JAMS | Zanjani et al.(2016) built a model of how a user's decisional procrastination (e.g., avoiding work) and tendency towards experimentation affect their online procrastination, internet "flow", and purchase behavior. |
| JAIS | Arbore et al. (2014) built a model of the antecedents of technology acceptance, which include perceived ease of use, usefulness, enjoyment and costs, along with social influence and measures of self-identity, driven by need for uniqueness and status gains. |
| JAIS | Goh, Tan, and Teo (2015) created a stated choice analysis experiment of consumer preferences for group choice institutions (GPI), modeling the type of order (best/final price), information cues, and the GPI risk profile. |
| MKSC | Häubl and Trifts (2000) examined consumer online decision making when exposed to recommendation agents and a comparison matrix. Both these methods reduced the size and increased the quality of consumer decision sets. |

| Journal | Description |
|---|---|
| MKSC | De Bruyn et al. 2008) built a conjoint analysis-based tool that finds customer preferences from prior purchases, psychographics, and demographics, and uses these preferences to make product recommendations. |
| MISQ | Duan et al. (2009) examined software download adoption behavior of popular products due to informational cascades from previous users and network externalities. |
| MISQ | Kim and Son (2009) described two interlinked mechanisms for post-website adoption consumer behavior, dedication, where perceived benefits affect loyalty and usage intention/WOM, and service/website investment, which affects switching costs and WTP/loyalty. |
| MISQ | Wang and Benbasat (2009) examined how different interactive decision aid (IDA) strategies (additive-compensatory, elimination, and hybrid) impact perceived advice quality, restrictiveness, cognitive effort, and intention to use. |
| MISQ | Kim (2009) extended the integrative framework of technology use into a panel data setting, where the determinants of technology usage (reason-oriented action, sequential updating, feedback, & habit) have proximal and distal effects. |
| MISQ | Dou et al. (2010) examined how priming (setting a desired product quality) moderates the effect of evaluation of an unknown brand positioned before well regarded brands in a search engine. |
| MISQ | Venkatesh, Thong, and Xu (2012) modified the UTAUT (unified theory of acceptance and use of technology) theory of user acceptance by incorporating additional consumer antecedents of behavioral intention (hedonic motivation, price value, habit) and additional demographic moderators. |

**Table 8. Cluster Three: Services & Loyalty**

| Journal | Description |
|---|---|
| ISR | Cenfetelli et al. (2008) examined how supporting services functionality (e.g., online reviews, recommendations) and service quality affect perceived usefulness and satisfaction in a B2B e-commerce. |
| ISR | Ba et al. (2010) built a game theory model to examine the optimal mix of investment in digital system services with human-based services. |
| ISR | Sun et al. (2012), in an IT services context, examined how different aspects of social capital (structural, cognitive, & relational) affect service quality and user satisfaction. |
| ISR | Xu, Benbasat, et al. (2014) examined how technologies with service provider variability compare with technologies that incorporate both service provider and user response variability with respect to efficiency and personalization. |
| ISR: | Xu, Thong, et al. (2014) built a model to show how service innovation (service + technology leadership and personalization) affects brand equity and loyalty. |
| ISR: | Hong and Pavlou (2017) built a model of how a range of cultural and geographical variables, along with provider information, impact selection of IT service providers. |
| JMIS | Grover et al. (1996) examined how service quality and the level of partnership affect the success of different types and extents of IT outsourcing. |
| JMIS | In a banking context and taking a service perspective, Tallon (2010) examined how business strategy and IT strategy align to affect firm performance. |
| JMIS | Benlian et al. (2011) extended previous marketing work on SERVQUAL to examine service quality in "software as a service" contexts. |
| JMIS | Benlian (2013) built a model of how perceptual congruence between IT professionals and users interacts with different service quality dimensions to drive customer satisfaction. |
| JMIS | Akçura and Ozdemi (2017) built a game theory model that described examines the decision on whether and how to offer online services in addition to face-to-face services. |

| Journal | Description |
| --- | --- |
| JAMS | Nysveen et al. (2005) built a model analyzing how different perceived service characteristics affect attitude and intension to use mobile services. |
| JAMS | Deleon and Chatterjee (2017), in a B2B context, built a structural model to examine how a seller's hard and soft service resources combine with factors related to buyer technology assimilation to affect relationship satisfaction. |
| JAIS | Xu et al. (2011) built a model for online firms of how perceived consumer sacrifice (e.g., time spent in enabling service) and perceived service quality affect service outcomes and customer loyalty. |
| MISQ | Pitt et al.(1995) examined the suitability of the SERVQAL scale for IT services and found it suitable, with some timitations (e.g., tangibles construct). |
| MISQ | Kettinger and Lee (1997) compared and contrasted the utility of different measures of service quality for IT services, including SERVQUAL, SERVPERF, SERVQUAL+ and IT-adapted versions of these scales. |
| MISQ | Susarla et al. (2003) built a model of application service provision, where prior experience, expectations, and disconfirmation affect perceived performance and satisfaction. |
| MISQ | Hsieh et al. (2012) built a model of how satisfaction with employee mandated CRM, embodied service knowledge, and job dedication all impact employee service quality and thus customer satisfaction. |
| MISQ | Tan et al. (2013) built a model of how in an e-government context, different aspects of service content quality and IT service delivery impact service quality. |
| MISQ | Setia et al. (2013) built a model showing how information quality moderated by process sophistication affects customer orientation, response capabilities, and service performance. |
| MISQ | Xu et al. (2013) built the 3Q model of e-services, which defines how system, information, and service quality affect satisfaction with these attributes, which then affects behavioral beliefs/attitudes to a service. |
| MISQ | Srivastava and Shainesh (2015) utilized service dominant logic and a healthcare case study to examine how information and communication technologies could be used to help bridge the service gap for disadvantaged groups in society. |
| MISQ | Tan et al. (2016) built a model of how different aspects of e-commerce system failure (informational, functional, and system) affect the disconfirmed expectancy. |

**Table 9. Cluster Four: Online Reviews**

| Journal | Description |
| --- | --- |
| ISR | Kumar and Benbasat (2006) ran a behavioral experiment in which they showed that incorporating consumer reviews and product recommendations improved the transactional and social aspects of the e-commerce experience. |
| ISR | Lu et al. (2013), using data from a Chinese e-commerce provider, showed that online review volume and valence along with coupon promotions positively impacted sales, with a negative interaction between review valence and coupons. |
| ISR | Kwark et al. (2014) built a game-theoretic model of the consequences of online reviews for retailers and manufacturers, with an emphasis on how reviews affect competition and consumers' utility functions. |
| ISR | Yin et al. (2016) found that consumers utilize confirmation bias when interpreting online reviews and find positive (negative) reviews more helpful when the average review scores are high (low). |
| ISR | Ho et al. (2017) examined disconfirmation in online reviews and found that reviewers who experienced disconfirmation between their experience and current aggregate review score were more likely to post a review. |
| JMIS | Dellarocas et al. (2010) described how for movies, consumers are motivated to review very popular or unpopular items, resulting in a u-shaped relationship between movie revenue and propensity to review. |
| JMIS | Li et al. (2011) built a game-theoretic model that analyzes how the informativeness of reviews affects profits under price competition and consumer switching. |
| JMIS | Benlian et al. (2012) contrasted provider recommendations (PR) and consumer reviews (CR), finding that PRs were perceived to be more useful and easier to use, but CRs had higher trust and greater affective qualities. |
| JMIS | Jensen et al. (2013) ran a behavioral experiment to find antecedents of review credibility, including lexical complexity, two-sidedness, and affect intensity. |
| JMIS | Huang et al. (2013) found that if the review type and product type matched (attribute-based □ search product, experiential □experience product) then the review was easier to understand and helpful. |
| JMIS | Ma et al. (2013) ran a longitudinal analysis how reviewer characteristics and prior reviews affect current reviews, e.g., reviewers with low experience and geographical mobility are more swayed by prior reviews. |
| JMIS | Li (2016) found that offering promotion deals (e.g., Groupon) affected review scores, moderated by previous review volume/valence, e.g., a company with a few poor reviews could increase reputation by offering promotions. |
| JMIS | Zhou and Duan (2016), using data from the CNET download site, found that professional reviews spur an increased volume of user reviews, which can then in turn lead to more purchases. |
| JM | Ludwig et al. (2003) found that for product conversion rate increases are driven by a nonlinear damped increase in positive affective content and by congruence in linguistic style between reviews and leaders. |
| JAIS: | Cheung et al. (2012) found that central (argument quality) and peripheral (source credibility, review consistency, and review sidedness) factors influence review credibility. |
| JAIS | Hong et al. (2016) found that cultural behavior affected reviewing behavior, with reviews from collectivist cultures less likely to deviate from existing reviews or show strong emotions. |
| MKSC | Godes and Silva (2012) built a temporal model of online ratings, looking at time and order effects, and examining antecedents and moderators (e.g., review homogeneity) of declining review scores. |
| MISQ | Mudambi and Schuff (2010) examined the effects of review rating, product type, and review depth on helpfulness. Generally, moderate ratings were the most helpful, though this was moderated by product type (search vs. experiential). |
| MISQ | Jabr and Zheng (2014) examined the combined impact of online reviews, opinion leadership, and product referrals, noting that the network centrality of a product in the referral space improves its competitiveness vis-à-vis other products. |
| MISQ | Shen et al. (2015) examined how reviewers gain attention. For Amazon, there was a trade-off, with more sales, less existing reviews, and higher deviation from existing reviews giving more (+ve and -ve) attention. |
| MISQ | Gao et al. (2015), for physician reviews, compared online reviews with offline population perceptions. The measures are correlated, but online reviews are mostly for physicians with higher offline perceptions and skew higher. |
| MISQ | Hu et al. (2017) built an analytical model of self-selection bias in online reviewing behavior, modeling both acquisition bias (towards favorable reviews) and underreporting bias (towards extreme reviews). |
| MISQ | Li et al. (2017) built a model of product review consumption using foraging theory and found that consumers first use attribute-oriented reviews followed by usage-oriented reviews. |
| MISQ: | Li (2017) built a game theoretic model examining profit under constraints of reveling or not revealing product review scores, product similarity with competitiors, and the heterogeneity of consumer valuation of the product. |

**Table 10. Cluster Five: Social Media and UGC (User-Generated Content)**

| Journal | Description |
|---|---|
| ISR | Dou, Niculescu, and Wu (2013) built a game theoretic network effects model to examine the interaction of different social media features with network seeding and pricing strategies. |
| ISR | Luo et al. (2013) built a model showing that social media metrics, such as blog sentiment and review ratings predict future firm equity value. |
| ISR | Goh et al. (2013) used a Facebook brand community dataset along with consumer transactions to show that the effects of UGC are greater than the effects of marketer-generated content (MGC) on brand purchases. |
| ISR | Zeng and Wei (2013) examined how as users build social ties in a social network, their posting behavior becomes more similar to the users with whom they share social ties. |
| ISR | Rishika et al. (2013) built a model of how a customer's social media participation affects their purchases, shopping visits, and overall firm relationships. |
| ISR | Ray et al. (2014) built a model of online community engagement that shows how knowledge self-efficacy, self-identity verification, and community identification affect community engagement and satisfaction, which leads to more contributions and more positive WOM. |
| ISR | Dewan et al. (2017) built a model of social influence for online music that incorporates both overall popularity/favorites and specific favorites from social network friends. |
| JMIS | Porter et al. (2013) contrasted consumer- and firm-sponsored online communities, by examining how sponsor effects (e.g., content embeddedness, interaction) affect trust constructs and behavioral outcomes. |
| JMIS | Luo and Zhang (2013) built a lagged autoregressive model of how online buzz, site traffic, and firm performance are related, using data from IT hardware and software vendors. |
| JMIS | Kuan et al. (2014) built a model of how in a social network, "buying" and "liking" deal information affect attitude, intention, and emotions (measured using EEG equipment). |
| JMIS | Chen et al. (2014) split users' social network behavior into content creation, content transmission, and individual and group relationship contributions, and then looked at the links between these constructs. |
| JMIS | Using Facebook business fan pages, Xie and Lee (2015) built a model of how earned social media and owned social media work together to impact consumer propensity to buy and with in-store promotions affect actual offline purchases. |
| JMIS: | Lee et al. (2015) built a model of how Facebook likes, along with other variables, such as price, discount, online review scores, and available days affect sales of e-commerce deals. |
| JMIS | Susarla et al. (2016) built a probit model of online video views using a range of channel characteristics, video characteristics and also a range of influence and network measures. |
| JMIS | Kuem et al. (2017) developed a tripartite model of participation in social networking sites. Here the antecedents of satisfaction, past investment, and social support inform commitment and participation. |
| JMIS | Pan et al. (2017) built a model of how a social network user's social and relational identity, moderated by inertia, affects use behavior (varied and reinforced). |
| JMIS | Gunarathne et al. (2017) examined the antecedents (including social media influence, repeated complaining, and process/outcome related issues) of social media complaint resolution satisfaction. |
| JAMS | Chung et al. (2016) built a predictive model for consumption personalized mobile content that incorporates adaption and social network information. |
| MKSC | Ransbotham et al. (2012) created a network effects model, which shows that the popularity of UGC is predicted by both the content and the network embeddedness and centrality of the contributor. |
| MISQ. | Animesh et al. (2011) examined how environmental stimuli, including technological and spatial effects, affect a user's experience in a virtual community (telepresence, flow, and social network effects), which in turn affects intention to purchase |

| Journal | Description |
|---|---|
| MISQ | Oestreicher-Singer and Zalmanson (2013) showed that as consumers become more engaged and participate more in online communities, they are more prepared to purchase premium services. |
| MISQ | Tsai and Bagozzi (2014) built a model of contribution for virtual communities that includes a range of antecedents of contribution intention and quality, including, subjective norms, group norms, social identity, attitudes, and emotions. |
| MISQ | Oh et al. (2016) showed that adding a paywall on a news site has a strong impact on social media WOM, particularly for popular articles, and lengthens the tail of the WOM distribution. |
| MISQ | Zhang et al. (2016) utilized network analysis and community detection on Facebook data to find brand networks consisting of brands related to a focal brand and sentiment analysis to recommend brands to users. |
| MISQ | Geva et al. (2017) examined how adding Google Trends search data to social media data (sentiment and mentions) improved prediction of purchase volume for car brands. |
| MISQ | Luo et al. (2017) built an autoregressive model that shows that expert blog sentiment in the technology industry is a leading indicator of consumer brand perceptions. |

**Table 11. Cluster Six: Supply Chain Management/Business-to-Business**

| Journal | Description |
|---|---|
| ISR | Saraf et al. (2007) showed how IS flexibility leads to IS integration and knowledge sharing between customers and channel partners, which leads to improved performance. |
| ISR | Bharadwaj et al. (2007) built a model of how manufacturing-marketing, manufacturing-IS (leading to integrated IS capability), and manufacturing-supply chain coordination work together to impact manufacturing performance. |
| ISR | Chatterjee and Ravichandran (2013) examined how in interorganizational information systems (IOS) implementations, the criticality and replaceability of resources influence the level of financial, and transactional IOS governance. |
| ISR | Dong et al. (2017) showed that lower institutional distance and IOS adaptability lead to improved IOS-enabled knowledge sharing between partners, which leads to improved performance. |
| JMIS | Son et al. (2005) examined how measures of corporate trust, uncertainty, and asset specificity increase cooperation between companies, which increases EDI (electronic data interchange) usage. |
| JMIS | Premkumar et al. (2005) utilized a theories of information processing to how the match between information processing needs and capability affect firm performance. |
| JMIS | Patnayakuni et al. (2006) built a model of how relational orientation (asset specificity, interactions, and long-term orientation) affect information flow integration across a B2B supply chain. |
| JMIS | Grover and Saeed (2007) built a model that explains IOS integration in terms of the information sharing environment, and demand, product, and market characteristics. |
| JAMS | Kim et al. (2006) showed that for a channel partnership, innovation drives integration, coordination, and information exchange, which leads to improved market performance. |
| JAMS | Davis and Golicic (2010) built a model showing how market-based IT competence creates market information, which leads to competitive advantage. |
| JAMS | Davis-Sramek et al. (2010) examine how environmental uncertainty moderates the relationships between both e-commerce integration and analytic IT capacity and performance. |
| JAMS | Nakata et al. (2011) built a model of IS/marketing functional integration and performance, where variables such as IT strategic intent, customer orientation, and marketing-IS structure are used to measure this integration. |

| Journal | Description |
|---|---|
| JAIS | Grover et al. (2002) examined how transaction costs and IT variables (monitoring, automation, and sharing) affects relationism between B2B buyers and suppliers. |
| MISQ | Christiaanse and Venkatraman (2002), in the context of B2B relationships between travel agents and airlines, examined how exploiting expertise in specific information systems improves channel integration. |
| MISQ | Goo et al. (2009) examined have different elements of formal B2B contracts affect relational elements, such as norms, conflict resolution, trust, and commitment. |
| MISQ | Klein and Rai (2009) built a structural model that incorporates trust, dependence, IT customization, bi-directional information flows between buyer and supply and relationship performance. |
| MISQ | Rai et al. (2012) build a framework that examines how interfirm communications and interfirm IT capabilities affect loyalty and share of wallet (proportion of buyer's purchases going to supplier). |
| MISQ | Wang et al. (2013) built a model of how a buyer's use of informational resources and normative contracts with a supplier affect a supplier's relation specific investments and relationship/modification flexibility. |

**Table 12. Cluster Seven: Information Privacy**

| Journal | Description |
|---|---|
| JCR | John et al. (2011) examine how indirect/direct messaging and the professionalism of an online survey affect the propensity of people to admit to risky, socially undesirable, or illegal behavior. |
| JMIS | Chellappa and Shivendu (2008) looked at the trade-offs between privacy and convenience of personalization services under different regulatory regimes (e.g., restricting use of cookies etc. and allowing the use of marketing information/cookies). |
| JMIS | Mai et al. (2010) showed that vendors displaying third party privacy seals or certificates can charge a higher price premium. |
| JMIS | Choi et al. (2006) used a service recovery perspective to examine how a firm's recovery strategy from privacy breaches affects consumer word of mouth and brand switching. |
| JAIS | Greenway and Chan (2005) examined how theories from the resource-based view and institution approach to the firm explain information privacy behaviors. |
| JAIS | Van Slyke et al. (2006) showed that consumer information privacy concerns affect willingness to purchase, a relationship which is mediated by trust and risk perceptions. |
| MISQ | Hui et al/ (2007) showed that privacy statements (+), monetary incentives (+), and the amount of information requested (-) affect customers' willingness to provide private information in an e-commerce setting. |
| MISQ | Son and Kim (2008) examined antecedents (privacy concerns, perceived justice, societal benefits) of different types of responses to information privacy issues (information provision, private/public action) |
| MISQ: | Angst and Agarwal (2009) used the ELM theory of attitude formation to show how messaging aspects of an appeal to opt-in to e-health records affect opt-in intention and assuage privacy concerns. |
| MISQ | Anderson and Agarwal (2010) built a model that relates security concerns, perceived user effectiveness, and security behavior self-efficacy to attitudes and intentions to perform security related behavior and then built a model to find the most effective message mix to change attitudes. |
| MISQ | Lee et al. (2011) created a game theoretic model, which explores privacy protection choices in the context of consumer personalization and segmentation. |
| MISQ | Koh et al. (2017) created a game theory model of how voluntary customer profiling improves social warfare and aggregate consumer surplus. |

| MISQ: | Lee et al. (2012) showed how the use of a filler interface with text and image components while a user waits for an e-commerce task to complete (e.g., a ticket purchase) affects a user's cognitive absorption, perceived wait time, website appraisal, and use intentions. |
|---|---|
| MISQ | Ou et al. (2014) built a model of how in China effective use of engagement tools (instant messenger, message box, feedback system) can increase customer perceptions of interactivity and presence, which leads to "Swift Guanzi", trust, and repurchases. |
| MISQ | Fang et al. (2014) examined how the perceived effectiveness of e-commerce governance and institutional mechanisms moderates the relationships between vendor satisfaction and vendor trust (+) and the relationship between vendor trust and repurchase intent (-). |